\documentclass[conference]{IEEEtran}
\IEEEoverridecommandlockouts
\usepackage{cite}
\usepackage{amsmath,amssymb,amsfonts}
\usepackage{graphicx}
\usepackage{textcomp}
\usepackage{xcolor}
\usepackage[hyphens]{url}  % DO NOT CHANGE THIS
\def\BibTeX{{\rm B\kern-.05em{\sc i\kern-.025em b}\kern-.08em
    T\kern-.1667em\lower.7ex\hbox{E}\kern-.125emX}}

\usepackage{amsmath}
\usepackage{amsfonts}
\usepackage{xcolor}
\usepackage[ruled,linesnumbered]{algorithm2e}
\usepackage{algpseudocode}
\usepackage{booktabs}
\usepackage{multirow}
\usepackage{rotating}
\usepackage{graphicx}
\usepackage{adjustbox}
\usepackage{pifont}
\usepackage{hyperref}

\newcommand{\figureBaseSkip}{\vspace{-2em}}

\begin{document}

\title{Modular Retrieval-Augmented Generalization for Human Action Recognition}

% \author{Anonymous ICME submission}

\author{
\IEEEauthorblockN{Peng Liao\IEEEauthorrefmark{2},
Shangsong Liang\IEEEauthorrefmark{2}\IEEEauthorrefmark{1},
Lin Chen\IEEEauthorrefmark{4}\IEEEauthorrefmark{1}
\thanks{\IEEEauthorrefmark{1}Corresponding authors: Shangsong Liang and Lin Chen.},
Peijia Zheng\IEEEauthorrefmark{2}}

% \IEEEauthorblockA{\IEEEauthorrefmark{2}School of Computer Science and Engineering, Sun Yat-sen University}
% \IEEEauthorblockA{\IEEEauthorrefmark{4}Engineering Research Centre of Applied Technology on Machine Translation and Artificial Intelligence, Macao Polytechnic University}
\IEEEauthorblockA{
\IEEEauthorrefmark{2}School of Computer Science and Engineering, Sun Yat-sen University\\
\IEEEauthorrefmark{4}Engineering Research Centre of Applied Technology on Machine Translation\\
and Artificial Intelligence, Macao Polytechnic University
}
% \IEEEauthorblockA{\IEEEauthorrefmark{4}CAI, Macao Polytechnic University}
\IEEEauthorblockA{Emails: liavonpenn@gmail.com, liangshangsong@gmail.com, lchen@mpu.edu.mo}
}

\maketitle

\begin{abstract}
Inertial Measurement Unit (IMU)-based Human Activity Recognition (HAR) aims to interpret and classify user behaviors from temporal motion signals.
Recently, deep learning frameworks have advanced this task by learning and extracting discriminative spatiotemporal representations, significantly improving recognition performance.
However, IMU-based HAR still faces several critical challenges, particularly limited training samples and static knowledge utilization, both of which severely hinder its large-scale deployment. In this paper, we introduce MoRA, the first \textbf{R}etrieval-\textbf{A}ugmented \textbf{Mo}dule specifically designed for motion series. It can be flexibly integrated into any existing HAR model, enhancing recognition performance while maintaining inference efficiency. To address issues such as information redundancy in retrieval results and rigid fusion strategies, we propose an uncertainty-adaptive fusion unit within MoRA. This unit leverages previous physical knowledge from IMU signals to dynamically adjust the fusion strategy between original outputs and retrieved information, enabling more robust recognition. Extensive experiments on ten real-world datasets demonstrate that MoRA significantly improves the performance of existing IMU-based HAR models, consistently delivering stable and effective gains.
The implementation of MoRA is publicly available at \url{https://github.com/liavonpenn/mora}.
\end{abstract}


%% ============================================================================ %%
\section{Introduction}
\label{sec:intro}

Human Activity Recognition (HAR) aims to automatically infer specific user actions or activity types from collected sensor data~\cite{liao2025wireless}. Compared to camera-based vision methods~\cite{ohnishi2016recognizing}, using Inertial Measurement Unit (IMU) sensors for HAR offers lower power consumption and better privacy. These sensors continuously collect tri-axial acceleration and angular velocity signals to produce structured time series data, providing critical support for HAR applications~\cite{zhang2022deep} in health monitoring, motion analysis, and virtual reality.

IMU-based HAR has advanced from traditional feature engineering to deep learning–based frameworks that significantly improve recognition accuracy and generalization capability~\cite{xu2023practically, zhang2024unimts}. 
% Recently, the introduction of large-scale models~\cite{girdhar2023imagebind, han2024onellm} has further extended HAR beyond static activity classification toward a more comprehensive understanding of human behavior and context awareness. 
Recently, large-scale cross-modal models~\cite{girdhar2023imagebind, han2024onellm} have extended HAR beyond static activity classification, enabling comprehensive cross-modal understanding of human behavior and context awareness.
However, performance degradation remains a major challenge in real-world deployments. This is primarily due to the gap between the controlled conditions of HAR model pre-training and the real-world deployment environments, particularly in terms of user behavior (e.g., movement speed and amplitude) and device-specific factors (e.g., hardware model and placement). Collecting large-scale personalized data is effective but labor-intensive, and thus impractical for widespread deployment.

\begin{figure}[t]
\centering
\includegraphics[width=0.9\columnwidth]{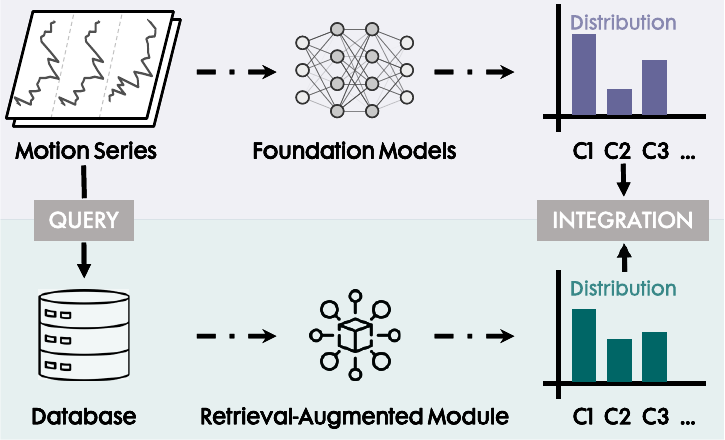}
\vspace{-0.5em}
\caption{Overview of the MoRA.}
\label{fig1}
\vspace{-1.5em}
\end{figure}

% To address performance bottlenecks from limited training data and static knowledge utilization, we develop

Motivated by the limitations of scarce training data and static knowledge utilization, we aim to introduce a retrieval‑augmented mechanism.
% ~\cite{fan2024survey}. 
The core idea is to dynamically retrieve relevant information from external or expanded databases during training or inference, and integrate it into the model as additional context to improve performance. Although retrieval augmentation has achieved significant success in fields such as computer vision and natural language processing, it remains largely unexplored in IMU-based HAR. Applying existing methods~\cite{liu2024retrieval} directly to this domain presents two primary challenges: \emph{Balancing real-time performance and discriminability.} Given the constraints of computational resources and response latency in deployment environments, it is critical to construct a semantic embedding space that effectively balances efficiency and discriminative power for fast retrieval. \emph{ 
Integrating dynamicity with generalization.} IMU data exhibit strong temporal dynamics and individual variability, often accentuated by domain shift, which requires robust mechanisms to leverage retrieved knowledge effectively.

To tackle the aforementioned challenges, we propose MoRA, a novel \textbf{R}etrieval-\textbf{A}ugmented \textbf{Mo}dule for IMU-based HAR models, as illustrated in Figure~\ref{fig1}. In the offline construction phase, historical IMU series are encoded into low-dimensional embeddings using a pre-trained model and stored in a database with their corresponding labels. 
In the online retrieval phase, 
MoRA retrieves similar motion patterns with millisecond-level latency.
In the decision fusion phase, MoRA generates a reference probability distribution over classes based on the retrieved information and integrates it with the prediction from the base model. As IMU data in real-world scenarios often suffer from significant distribution shifts, MoRA integrates an uncertainty-adaptive fusion unit to dynamically adjust the decision-fusion strategy.

Our contributions can be summarized as follows: 
\ding{172} \textbf{We propose a novel HAR framework} incorporating MoRA, a retrieval-augmented module specifically designed for IMU data. MoRA features a highly modular architecture that can be seamlessly integrated into any existing IMU-based model. Without requiring additional training, it provides discriminative, retrieval-based knowledge during inference, thereby significantly enhancing cross-domain generalization.
\ding{173} \textbf{We introduce an uncertainty-adaptive fusion unit} in MoRA that leverages the inherent physical properties of motion data to adjust the fusion strategy dynamically. This unit improves MoRA's performance while effectively reducing retrieval overhead. 
\ding{174} \textbf{We perform extensive experiments} on ten real-world datasets by integrating MoRA into multiple state-of-the-art IMU models. Experimental results demonstrate that MoRA consistently achieves significant and stable improvements across various models and task settings.

%% ============================================================================ %%

\section{Methodology}

\begin{figure*}[htbp]
\centering
\includegraphics[width=0.9\textwidth]{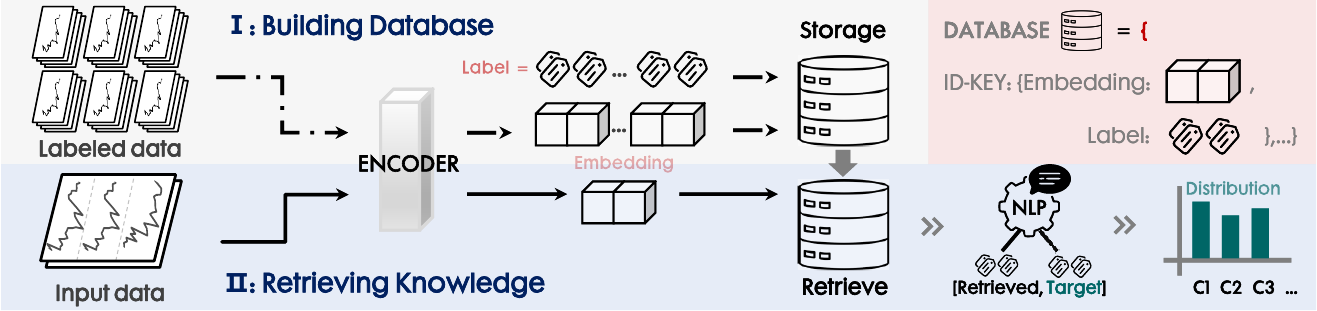} 
\caption{Workflow of the MoRA.}
\label{fig2}
\vspace{-1.5em}
\end{figure*}

We propose MoRA, a \textbf{Mo}dular \textbf{R}etrieval-\textbf{A}ugmented framework designed to enhance IMU-based HAR models. As illustrated in Figure~\ref{fig2}, MoRA operates in two stages: \textbf{(I)} constructing a multimodal knowledge base from large-scale historical data by storing embeddings with their semantic labels, and \textbf{(II)} retrieving relevant knowledge via embedding similarity to refine model predictions using natural language processing.
MoRA pseudocode is provided in Appendix~\ref{sec:mora_code}.

\subsection{Task formulation}

This paper focuses on time-series classification for HAR using data collected from body-worn IMU sensors. Given a dataset $\mathcal{D} = \{(x_i, y_i)\}_{i=1}^{n}$, where $x_i \in \mathcal{X}$ denotes an input instance and $y_i \in \mathcal{Y}$ its corresponding class label, the HAR model $f_\theta: \mathcal{X} \rightarrow \mathcal{Y}$ aims to predict labels for unseen samples. To further improve the performance of $f_\theta$, we propose a retrieval-augmented module that can be seamlessly integrated into existing models without affecting inference efficiency.

Specifically, we construct a multimodal database using historical samples $x$ and their labels $y$. To balance memory and performance, embeddings $z$ are first extracted from $x$ using a pre-trained encoder $\mathcal{F}$, and then organized into a key–value database $\mathcal{B} = \{(z_j^{\text{db}}, y_j^{\text{db}})\}_{j=1}^{m}$, where embedding vectors serve as retrieval keys and labels as corresponding values. This design enables language-based reasoning and supports dynamic updates. For a query sample $x_q$, the system retrieves the top-$k$ most similar embeddings from $\mathcal{B}$ and collects their associated labels $\{y_1^{\text{db}},\ldots,y_k^{\text{db}}\}$. 
We then apply natural language modeling~\cite{radford2021learning} to these retrieved labels to produce a class-wise semantic similarity distribution $\mathcal{C}_{\text{rag}}(y \mid x_q, \mathcal{B})$, which estimates the likelihood that $x_q$ belongs to each class.
Meanwhile, $f_\theta$ generates an initial distribution $\mathcal{C}_{\text{ref}}(y \mid x_q)$.  To effectively integrate model predictions with external prior knowledge, we perform a weighted fusion of the two distributions using a fusion ratio $\alpha \in [0, 1]$, resulting in the final prediction distribution:
\begin{equation}
    \mathcal{C}_{\text{final}}(y \mid x_q) = \alpha \cdot \mathcal{C}_{\text{ref}}(y \mid x_q) + (1 - \alpha) \cdot \mathcal{C}_{\text{rag}}(y \mid x_q, \mathcal{B})\,,
\label{eq-1}
\end{equation}
where $\alpha$ can be dynamically adjusted within the proposed uncertainty-adaptive fusion unit in MoRA to enhance robustness and discriminative performance.

\subsection{Retrieval-Augmented Module}

\noindent \textbf{Step I: Building databases.}
To enhance the effectiveness of the retrieval-augmented module, it is essential to construct a high-quality retrieval database. A common approach is to use a large-scale, standardized, and widely adopted dataset as the foundation for building the retrieval library. However, in the domain of IMU-based HAR, although several public datasets~\cite{grauman2024ego} exist with considerable scale and multimodal sensor coverage, most of them are collected under controlled experimental conditions that differ significantly from real-world deployment scenarios~\cite{kong2019mmact} involving consumer devices. 

Therefore, in the practical deployment of MoRA, we choose to build a task-specific retrieval database using historical data from the target domain. Let the historical dataset be defined as $\mathcal{D}_{\text{hist}} = {(x_j, y_j)}_{j=1}^{m}$; we employ the pretrained encoder $\mathcal{F}$ to extract embeddings $z_j = \mathcal{F}(x_j)$ for each instance, thereby constructing a structured database $\mathcal{B} = {(z_j^{\text{db}}, y_j^{\text{db}})}_{j=1}^{m}$. This database is organized in a key-value format, where the key $z_j^{\text{db}}$ supports efficient similarity-based retrieval in the embedding space, and the value $y_j^{\text{db}}$ provides auxiliary information in the label space. Notably, the database construction is performed entirely offline and remains fixed during inference, thereby introducing no additional computational overhead at runtime.

\noindent \textbf{Step II: Retrieving knowledge.}
Given a query instance $x_q$, MoRA retrieves semantically similar samples from the database $\mathcal{B}$ to enhance prediction.  The input is first encoded by the same feature extractor used in Step I to obtain an embedding $z_q$, ensuring consistency between database construction and inference. Efficient retrieval is achieved using FAISS~\cite{johnson2019billion}, which identifies the top-$k$ nearest neighbors $\{z^{\text{db}}_1, ..., z^{\text{db}}_k\}$ and their associated labels $\{y^{\text{db}}_1, ..., y^{\text{db}}_k\}$ within millisecond-level latency.
To leverage the retrieved information, MoRA concatenates the label set into a unified text sequence and encodes it using a pretrained CLIP text encoder~\cite{radford2021learning} to obtain a semantic representation $t_q$. The semantic  similarities between $t_q$ and the class-wise text embeddings $\{t^1, t^2, \dots, t^k\}$ are then calculated to produce a cross-modal retrieval-based $\mathcal{C}_{\text{rag}}(y \mid x_q, \mathcal{B}) = \text{softmax}(\tau \cdot (t_q \cdot T^\top))$, where $\tau$ is a temperature parameter that controls the sharpness of the distribution.
Finally, this retrieval-based distribution is fused with the model’s intrinsic prediction $\mathcal{C}_{\text{ref}}(y \mid x_q)$ through a weighted combination, following Eq.~\ref{eq-1}. 
While a fixed fusion ratio $\alpha$ can balance the two sources of information, its static nature fails to adapt to varying input uncertainty and retrieval quality in real-world scenarios.

\subsection{Uncertainty-Adaptive Fusion Unit}

% To enhance decision-making under diverse conditions, it is desirable for MoRA to adjust the contribution of retrieved knowledge relative to model predictions on a per-instance basis. We therefore propose an uncertainty-adaptive fusion unit that provides MoRA with a flexible and dynamic fusion mechanism. The key idea is to adjust the fusion ratio adaptively based on instance-specific features, thereby improving robustness under real-world conditions. Concretely, we design a lightweight gating network that predicts an instance-specific $\alpha$ using a set of reliable features. Using features from a pre-trained encoder is straightforward but yields limited discriminative power, whereas auxiliary features extracted by lightweight encoders are often redundant and provide little additional information. Considering these factors, we adopt a set of physics-informed features (definitions provided in Appendix~\ref{feature}), which are concatenated, normalized, and passed through a fully connected layer to predict $\alpha$.

To enable robust decision-making under diverse conditions, MoRA incorporates an uncertainty-adaptive fusion unit that dynamically adjusts the contribution of retrieved knowledge relative to model predictions on a per-instance basis. The fusion ratio $\alpha$ is predicted by a lightweight gating network using a carefully selected set of physics-informed features (definitions in Appendix~\ref{feature}), which are concatenated, normalized, and processed through a fully connected layer. While features from pre-trained encoders are readily available, they offer limited discriminative power, and auxiliary features extracted by lightweight encoders often introduce redundancy or noise into the system. By leveraging physics-informed features, the gating network effectively captures informative, task-relevant cues, enabling context-aware and adaptive fusion that substantially enhances robustness and classification performance in complex, real-world HAR scenarios and environments.

To train the gating unit effectively, we define a composite objective with three loss components.
First, the classification loss $\mathcal{L}{_\text{cls}}$ enhances prediction discriminability. Second, the entropy-guided alignment loss $\mathcal{L}_{\text{align}}$ aligns the fusion rate $\alpha$ with model uncertainty. Third, the sparsity loss $\mathcal{L}_{\text{sparse}}$ penalizes over-reliance on fusion to improve interpretability and robustness.
Details of each loss function are provided in the Appendix ~\ref{loss}.
The training objective is a weighted sum:
\begin{equation}
    \mathcal{L}_{\text{total}} = \boldsymbol{\lambda}^\top \cdot [\mathcal{L}_{\text{cls}},\; \mathcal{L}_{\text{align}},\; \mathcal{L}_{\text{sparse}}], \quad \text{where } \boldsymbol{\lambda} \in \mathbb{R}^3
\end{equation}
where $\boldsymbol{\lambda} \in \mathbb{R}^3$ is a learnable vector that adapts the relative contribution of each term during training. 
% Notably, this training scheme updates only the gating unit, keeping the backbone model frozen to reduce computational cost and enhance convergence stability. The trained MoRA can generate an adaptive fusion ratio $\alpha$ for each input.
Notably, this training scheme updates only the lightweight gating unit, keeping the backbone HAR model frozen to significantly reduce computational cost and enhance convergence stability during training. Once trained, the MoRA framework can generate an adaptive, instance-specific fusion ratio $\alpha$ for each input, enabling context-aware and dynamically optimized integration of retrieved knowledge with model predictions.

%% ============================================================================ %%

\section{Experiment Setup}

\begin{table*}[htbp]
  \centering
  \caption{Retrieval-augmented inference without fine-tuning. Methods marked as “N/A (+X)” do not support zero-shot classification, where “+X” indicates the performance achieved with our framework. For zero-shot-capable methods, values in parentheses denote the performance change after adaptation.}
  % \fontsize{7.65pt}{8pt}\selectfont
  \label{tab:rai_no_ft}
  \begin{adjustbox}{max width=\textwidth}
  \vspace{-1em}
  \begin{tabular}{c|c|cccccccccc}
    \toprule

    Baselines & Mets & UCI-HAR & MotionSense & Shoaib & RealWorld & PAMAP & USC-HAD & WISDM & DSADS & UTD-MHAD & MMAct \\
    \midrule
    \multirow{2}{*}{TS2Vec} & Acc & N/A (+71.2) & N/A (+77.4) & N/A (+78.7) & N/A (+54.9) & N/A (+65.5) & N/A (+39.9) & N/A (+53.0) & N/A (+41.6) & N/A (+29.3) & N/A (+34.3) \\
    & F1 & N/A (+70.6) & N/A (+76.5) & N/A (+79.0) & N/A (+53.3) & N/A (+63.2) & N/A (+42.4) & N/A (+52.7) & N/A (+39.8) & N/A (+32.4) & N/A (+27.2) \\
    \midrule
    \multirow{2}{*}{UniHAR} & Acc & N/A (+71.6) & N/A (+69.9) & N/A (+73.8) & N/A (+54.6) & N/A (+59.9) & N/A (+41.4) & N/A (+48.8) & N/A (+39.1) & N/A (+24.0) & N/A (+30.6) \\
    & F1 & N/A (+70.7) & N/A (+69.3) & N/A (+73.3) & N/A (+53.8) & N/A (+59.3) & N/A (+42.8) & N/A (+48.3) & N/A (+37.5) & N/A (+27.7) & N/A (+23.4) \\
    \midrule
    \multirow{2}{*}{TSLANet} & Acc & N/A (+70.1) & N/A (+64.5) & N/A (+60.7) & N/A (+45.5) & N/A (+42.1) & N/A (+27.2) & N/A (+30.1) & N/A (+25.9) & N/A (+6.3) & N/A (+16.5) \\
    & F1 & N/A (+68.7) & N/A (+56.8) & N/A (+59.9) & N/A (+38.5) & N/A (+37.7) & N/A (+25.1) & N/A (+31.1) & N/A (+22.0) & N/A (+4.4) & N/A (+12.6) \\
    \midrule
    \multirow{2}{*}{Mantis} & Acc & \textbf{N/A (+89.1)} & \textbf{N/A (+96.6)} & \textbf{N/A (+93.3)} & N/A (+68.3) & N/A (+74.4) & \textbf{N/A (+59.2)} & \textbf{N/A (+69.1)} & N/A (+65.1) & N/A (+38.6) & N/A (+51.6) \\
    & F1 & \textbf{N/A (+88.9)} & \textbf{N/A (+95.9)} & \textbf{N/A (+93.2)} & N/A (+69.3) & N/A (+76.4) & \textbf{N/A (+62.6)} & \textbf{N/A (+69.2)} & N/A (+63.2) & N/A (+45.4) & N/A (+45.4) \\
    \midrule\midrule
    \multirow{2}{*}{IMU2CLIP} & Acc & 14.8 (+56.7) & 26.7 (+54.3) & 19.1 (+58.2) & 8.2 (+53.6) & 7.3 (+54.1) & 18.2 (+32.4) & 3.7 (+47.1) & 5.1 (+40.2) & 5.7 (+19.3) & 2.3 (+35.7) \\
    & F1 & 8.3 (+61.2) & 12.9 (+60.5) & 7.1 (+65.4) & 6.7 (+52.4) & 4.5 (+54.9) & 10.5 (+36.2) & 2.5 (+46.8) & 3.2 (+39.8) & 3.2 (+24.5) & 1.8 (+30.2) \\
    \midrule
    \multirow{2}{*}{UniMTS} & Acc & 35.2 (+52.1) & 45.2 (+50.8) & 63.6 (+27.2) & \textbf{43.6 (+24.8)} & \textbf{47.2 (+28.7)} & 30.5 (+16.6) & 27.8 (+32.2) & \textbf{37.5 (+48.3)} & \textbf{22.8 (+35.8)} & \textbf{10.2 (+43.2)} \\
    & F1 & 22.0 (+65.3) & 33.7 (+62.2) & 57.2 (+33.2) & \textbf{36.7 (+36.9)} & \textbf{43.6 (+35.7)} & 27.8 (+24.9) & 25.5 (+35.1) & \textbf{23.7 (+54.1)} & \textbf{18.5 (+37.0)} & \textbf{10.0 (+37.5)} \\
    \midrule
    \multirow{2}{*}{PRIMUS} & Acc & 12.4 (+54.5) & 28.4 (+52.7) & 17.4 (+55.9) & 6.1 (+50.8) & 5.0 (+52.3) & 15.6 (+29.9) & 1.3 (+44.8) & 2.9 (+38.4) & 3.5 (+17.0) & 1.0 (+33.4) \\
    & F1 & 5.9 (+59.0) & 15.1 (+62.8) & 5.8 (+67.7) & 5.0 (+50.7) & 2.2 (+56.3) & 7.8 (+33.7) & 1.0 (+44.0) & 1.6 (+37.6) & 1.1 (+22.3) & 0.5 (+28.6) \\
    \midrule\midrule
    \multirow{2}{*}{ImageBind} & Acc & 14.1 (+1.8) & 15.1 (+16.8) & 12.5 (+5.7) & 12.9 (+1.2) & 9.3 (+3.9) & 10.3 (+0.0) & 6.8 (+2.2) & 3.0 (+1.2) & 2.8 (+0.0) & 3.5 (+0.0) \\
    & F1 & 5.1 (+1.4) & 16.1 (+8.6) & 11.0 (+5.2) & 8.7 (+1.1) & 5.9 (+4.0) & 6.7 (+0.1) & 4.8 (+2.0) & 2.2 (+2.1) & 0.7 (+0.0) & 1.3 (+0.0) \\
    \midrule
    \multirow{2}{*}{OneLLM} & Acc & 15.0 (+63.7) & 25.8 (+48.8) & 14.3 (+64.1) & 16.6 (+39.2) & 10.7 (+54.3) & 6.8 (+26.9) & 7.3 (+47.9) & 5.4 (+41.4) & 4.4 (+19.5) & 4.3 (+28.9) \\
    & F1 & 13.4 (+64.9) & 11.3 (+60.9) & 9.4 (+68.6) & 6.5 (+47.0) & 5.6 (+60.1) & 4.7 (+29.9) & 3.5 (+51.2) & 2.6 (+42.4) & 1.4 (+25.6) & 2.0 (+23.8) \\
    \bottomrule
  \end{tabular}
  \end{adjustbox}
  \vspace{-1em}
\end{table*}

\subsection{Research Questions}
We conducted a series of experiments to systematically evaluate the proposed retrieval-augmented module by addressing six research questions: (RQ1) Can MoRA enhance the performance of existing IMU-based HAR models? (RQ2) How do different retrieval strategies affect its effectiveness? (RQ3) How does the label concatenation strategy influence fusion performance? (RQ4) How sensitive is MoRA to key hyperparameters? (RQ5) Can it generalize to unseen scenarios or new tasks? (RQ6) What is its computational overhead?

\subsection{Datasets and Baselines}

We comprehensively evaluated the proposed method on ten widely used benchmark datasets for IMU-based HAR.
These datasets include diverse daily activities with sensors placed on various body parts—head, chest, back, arms, wrists, and waist—reflecting realistic scenarios.
The selected datasets include UCI-HAR~\cite{anguita2013public}, MotionSense~\cite{malekzadeh2019mobile}, Shoaib~\cite{shoaib2014fusion}, RealWorld~\cite{sztyler2016body}, PAMAP~\cite{reiss2012introducing}, USC-HAD~\cite{zhang2012usc}, WISDM~\cite{weiss2019wisdm}, DSADS~\cite{altun2010comparative}, UTD-MHAD~\cite{chen2015utd}, and MMAct~\cite{kong2019mmact}. 

To evaluate MoRA’s effectiveness, we integrated it into several SOTA HAR models for validation. We selected ten widely used open-source models spanning three modeling paradigms: (a) unimodal methods, including TS2Vec~\cite{yue2022ts2vec}, UniHAR~\cite{xu2023practically}, TSLANet~\cite{eldele2024tslanet}, Mantis~\cite{feofanov2025mantis}, and TimeMixer~\cite{wang2024timemixer++}; (b) bimodal methods, including IMU2CLIP~\cite{moon2022imu2clip}, UniMTS~\cite{zhang2024unimts}, and PRIMUS~\cite{das2025primus}; and (c) multimodal methods, including ImageBind~\cite{girdhar2023imagebind} and OneLLM~\cite{han2024onellm}. 
For models with available pre-trained weights, we used the released versions; otherwise, we retrained them following the original settings to ensure fair evaluation. Further details are provided in Appendices~\ref{dataset}--\ref{baseline}.

\subsection{Settings}

All experiments were implemented in PyTorch. For SOTA methods without publicly available pretrained weights, encoders were retrained using the Ego4D dataset~\cite{grauman2022ego4d}. Data splits and preprocessing for downstream HAR datasets followed the UniMTS protocol. MoRA was evaluated using accuracy and macro-F1, with training samples used to build the retrieval database for inference. All models were trained on four NVIDIA RTX 4090 GPUs using AdamW with a learning rate of $1\times10^{-4}$. Key hyperparameters were set as follows: fusion ratio $\alpha = 0.5$, number of retrieval candidates $k = 5$, and temperature $\tau = 20$. Sensitivity analysis is presented later.

%% ============================================================================ %%

\section{Results and Analysis}

\subsection{Overall Performance}

\noindent \textbf{RQ1:} The effectiveness of integrating MoRA is assessed under three settings: i) inference without fine-tuning, ii) inference with fine-tuning, and iii) inference with full training.

\noindent \emph{\textbf{Inference without fine-tuning}}. 
We constructed the retrieval database using the training set of each target-domain dataset and deployed MoRA without additional label supervision. For several SOTA models that lack inherent zero-shot inference capability, the fusion ratio was fixed at $\alpha = 0$, relying solely on retrieval-augmented inference. The uncertainty-adaptive fusion unit was not employed during this evaluation. Table~\ref{tab:rai_no_ft} summarizes the performance of nine representative HAR models integrated with MoRA across ten benchmark datasets. Results show that MoRA significantly improves model capability and generalization. For instance, Mantis achieved accuracy gains of 89.1\% and 96.6\% on UCI-HAR and MotionSense, respectively (denoted as N/A (+X)), indicating that MoRA can endow non-zero-shot models with strong generalization performance comparable to that of fully trained counterparts. 
In contrast, large-scale multimodal models pre-trained on high-resolution sensory data performed poorly on low-frequency IMU signals. For example, OneLLM achieved only 15.0\% and 25.8\% accuracy on UCI-HAR and MotionSense, respectively. This suggests that the distribution mismatch between pre-training data and downstream domains substantially affects model transferability. Following the Wilcoxon signed-rank test, all improvements introduced by MoRA are statistically significant with p-values below 0.05. Overall, MoRA demonstrates strong generalization, robustness, and low deployment cost, making it an effective enhancement for real-world HAR systems.

\begin{figure}
    \centering
    \includegraphics[width=\columnwidth]{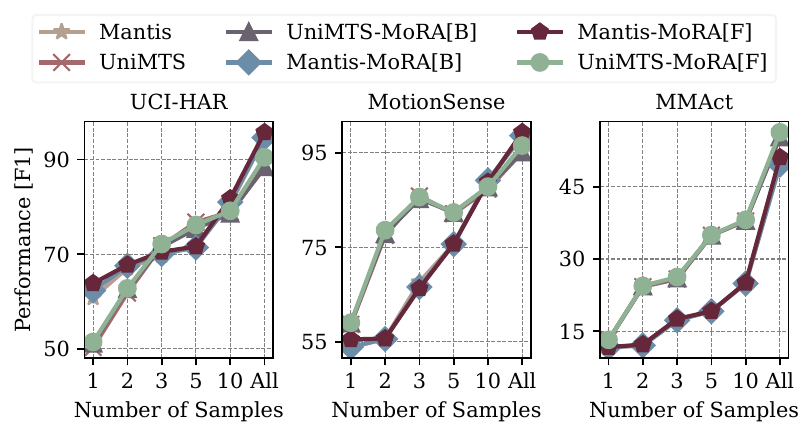}
    \figureBaseSkip
    % \vspace{-1.5em}
    \caption{Retrieval-augmented inference with fine-tuning. }
    \vspace{-2em}
    \label{fig:rai_ft}
\end{figure}

\noindent \emph{\textbf{Inference with fine-tuning}}.
We constructed training subsets containing 1, 2, 3, 5, 10, and all available samples per class. These subsets were used to build the retrieval database and fine-tune MoRA, as well as two representative SOTA baselines, Mantis and UniMTS, which demonstrated the best average performance in previous experiments. Due to space limitations, Figure 3 reports F1 scores on selected datasets; comprehensive results are provided in Appendix~\ref{tab:full_fine-tuning}. We report results for two MoRA variants: MoRA[B], the base version with static retrieval integration, and MoRA[F], the full version that incorporates the uncertainty-adaptive fusion mechanism. Experimental findings indicate that MoRA continues to offer performance gains even on top of fine-tuned models. Although the average improvement in F1 score is modest (1.5\%), MoRA achieves a notable 13.50\% average relative error rate reduction under full fine-tuning. In the MotionSense dataset, the error rate reduction reaches as high as 54.6\%.
Moreover, MoRA[F] requires only about one-seventh the training cost of standard fine-tuning, demonstrating its efficiency and practicality.
% Following the same Wilcoxon signed-rank test as in the prior setting, we validated MoRA's reliability and effectiveness.

\begin{figure}[htbp]
    \centering
    \vspace{-1em}
    \includegraphics[width=\columnwidth]{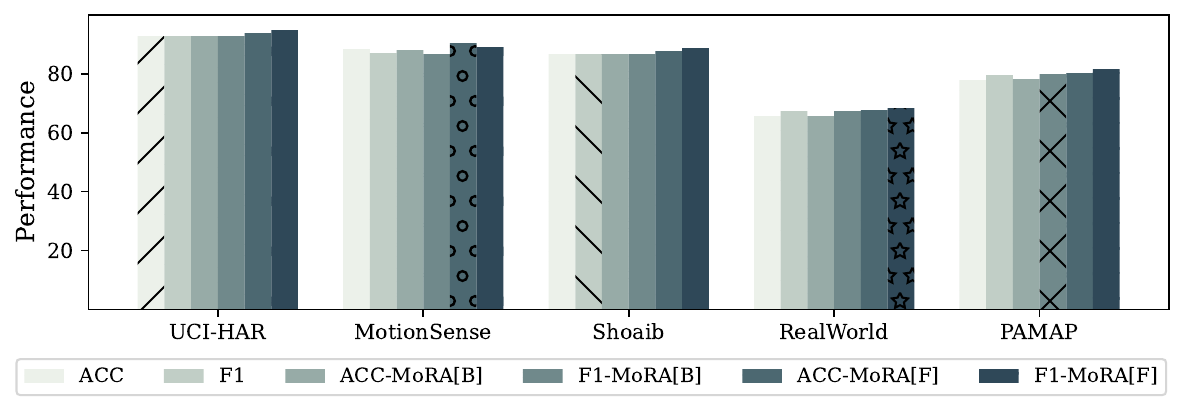}
    \figureBaseSkip
    % \vspace{-1.5em}
    \caption{Retrieval-augmented inference with full-training.}
    \vspace{-0.8em}
    \label{fig:rai_e2e}
\end{figure}

\noindent \emph{\textbf{Inference with full-training}}.
In the previous experiments, the encoder used as the feature extractor was a pretrained model, with no exposure to target domain data. Given that MoRA's retrieval-augmented mechanism is highly dependent on the semantic quality of the embedding representations, its performance is closely tied to the encoder's generalization capacity (see Appendix~\ref{eva_1} for further analysis). To examine MoRA's applicability, we selected TimeMixer as the backbone model and conducted full training from scratch, followed by integration of MoRA for evaluation. Figure 4 shows the performance across five datasets. We observed that integrating MoRA[B] yields classification accuracy and F1 scores comparable to the original TimeMixer, suggesting that the discriminative embedding space learned via end-to-end optimization may already be sufficiently robust, thereby diminishing the marginal utility of external semantic augmentation. In contrast, when MoRA[F] is incorporated, TimeMixer exhibits an average improvement of approximately 2.1\% in both accuracy and macro-F1 score. The improvement is statistically significant, as confirmed by a Wilcoxon signed-rank test.
% with a p-value = $1.3 \times 10^{-15}$
These findings demonstrate that MoRA is not only effective in settings with frozen pretrained encoders but also compatible with end-to-end training paradigms, offering stable performance gains.

\begin{table}[!tbp]
  \centering
  \caption{Influence of retrieval mechanism [F1]. $D$ denotes embedding dimension; “+” adds a linear mapping layer, and “*” indicates fine-tuning of the encoder.}
  \label{tab:rai_emd}
  \fontsize{7.5pt}{8pt}\selectfont
  \begin{adjustbox}{max width=\columnwidth}
  \vspace{-1em}
  \begin{tabular}{c|ccc|ccc|cc}
    \toprule

    Dataset & CCF & DTW & Pearson & D128 & D256 & D512 & D512+ & D512*\\
    \midrule
    USC-HAD & 58.45 & 60.05 & 59.26 & 59.13 & 59.76 & 59.51 & 60.95 & \textbf{64.4} \\
    \midrule
    WISDM & 70.66 & 70.79 & 70.6 & 71.22 & 71.56 & 73.71 & 73.76 & \textbf{80.13} \\
    \midrule
    DSADS & 92.19 & 92.24 & 91.34 & 94.27 & 93.67 & 93.76 & \textbf{94.46} & 92.8 \\
    \midrule
    UTD-MHAD & 66.59 & 65.18 & 67.59 & 66.09 & 69.79 & 70.09 & 70.77 & \textbf{85.03} \\
    \midrule
    MMAct & 55.51 & 52.47 & 55.39 & 56.54 & 56.04 & 56.28 & 58.79 & \textbf{71.76} \\
    \bottomrule
  \end{tabular}
  \end{adjustbox}
  \vspace{-1.8em}
\end{table}

\subsection{Sensitivity Analysis}

\begin{figure}[htbp]
    \centering
    \vspace{-1.5em}
    \includegraphics[width=\columnwidth]{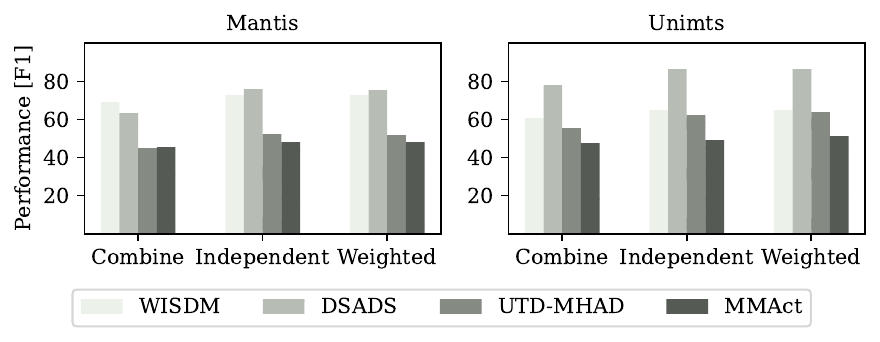}
    \figureBaseSkip
    % \vspace{-1.5em}
    \caption{Influence of label concatenation strategies.}
    \label{fig:rai_text}
    \vspace{-0.5em}
\end{figure}

\noindent \textbf{RQ2:}
To investigate the impact of different retrieval mechanisms on inference performance, we conducted a comparative study under the UniMTS framework; similar findings hold across other frameworks.
% ~\SL{(Applying to other frameworks essentially leads to the same conclusions)}.
Results are summarized in Table~\ref{tab:rai_emd}. Specifically, we evaluated two categories of retrieval strategies: (1) similarity-based retrieval, including cross-correlation function (CCF), dynamic time warping (DTW), and Pearson correlation, which operate directly on raw series; and (2) embedding-based retrieval, where representations are extracted using a pre-trained encoder. Since UniMTS generates high-dimensional embeddings (512D), we applied two strategies: principal component analysis (PCA) and a learnable linear layer. The experimental results demonstrate that embedding-based retrieval generally outperforms signal-based methods. However, when using raw, high-dimensional embeddings without projection, the performance advantage becomes marginal. Further improvements can be achieved by fine-tuning the encoder on the target dataset, thereby enhancing MoRA's effectiveness. Nevertheless, such gains come at the cost of additional computational resources, underscoring the trade-off between performance and efficiency in deployment.

\noindent \textbf{RQ3:}
To evaluate the impact of different information utilization strategies, we compared three schemes: \emph{Combine}, which concatenates all retrieved labels into a single text input; \emph{Independent}, which encodes each retrieved label separately and averages the scores; and \emph{Weighted}, which applies an exponential decay (rate = 0.9) based on retrieval rank. As shown in Figure~\ref{fig:rai_text}, using the Mantis model as a basis, the Independent strategy yields an average F1 improvement of approximately 6–7\% over Combine. However, this comes at the cost of a 3–5× increase in inference latency. 

\begin{figure}[htbp]
    \centering
    \vspace{-1em}
    \includegraphics[width=\columnwidth]{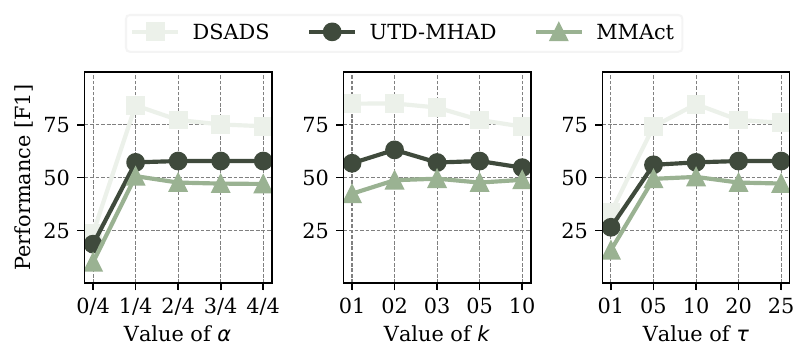}
    \figureBaseSkip
    % \vspace{-0.5em}
    \caption{Influence of hyperparameter choices.}
    \label{fig:para}
    \vspace{-0.5em}
\end{figure}

\noindent \textbf{RQ4:}
To evaluate MoRA's sensitivity to key hyperparameters, we conducted ablation studies on three factors: the fusion ratio $\alpha$, the number of retrieved candidates $k$, and the temperature $\tau$. The corresponding results are illustrated in Figure~\ref{fig:para}. We observed a clear performance gain as $\alpha$ increased from 0, confirming the effectiveness of integrating MoRA. On several datasets, the optimal F1 score was achieved around $\alpha=0.25$, after which performance tended to saturate or slightly fluctuate. A small $k$ yielded the best overall results, whereas larger $k$ values may introduce irrelevant information. In terms of $\tau$, the model exhibited stable behavior when $\tau$ was set to 10 or 20, indicating that moderate temperature scaling helps balance the sharpness of semantic similarity distributions. To fully exploit its potential in practical deployments, we recommend task-specific tuning of $\alpha$, $k$, and $\tau$ to optimize the synergy between retrieval-enhanced semantics and the primary model.

\subsection{Unseen Generalization}

\begin{figure}[htbp]
    \centering
    \vspace{-1em}
    \includegraphics[width=\columnwidth]{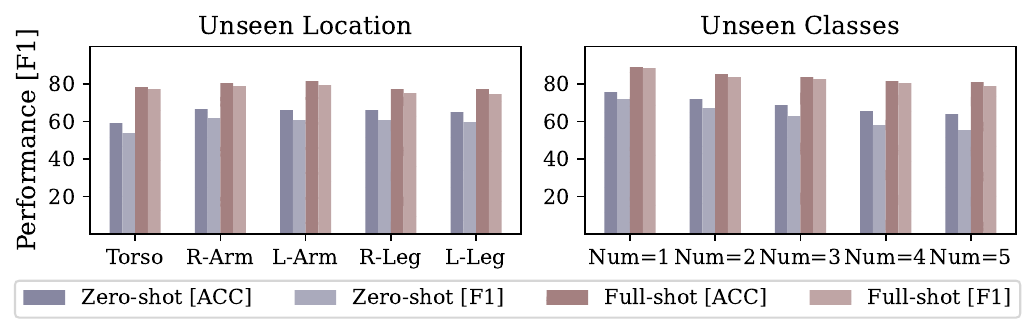}
    \figureBaseSkip
    \caption{Unseen scenarios. R/L denote `right’ and `left’.}
    \label{fig:unseen}
    \vspace{-0.5em}
\end{figure}

\noindent \textbf{RQ5:}
To evaluate MoRA’s generalization to unseen scenarios, we conduct experiments on the DSADS dataset, which features diverse sensor placements and a wide range of activity classes. We design two settings: location and classes. The performance of UniMTS enhanced with MoRA under both settings is reported in Fig~\ref{fig:unseen}. Results demonstrate that the Full-Shot condition yields overall higher performance. For instance, at the “Right Arm” position, the accuracy and F1 score differences between Full-Shot and Zero-Shot are only 1.46 and 5.51, respectively. This indicates that MoRA’s retrieval mechanism contributes to maintaining stable inference. In summary, MoRA not only enhances performance under standard training conditions but also exhibits strong generalization ability.

\subsection{Efficiency Analysis}

\noindent \textbf{RQ6:} MoRA has approximately 63M parameters, most of which are contained in a frozen text encoder, while the core module contains only about 4K trainable parameters. By storing information as a structured database instead of full historical data, memory usage is substantially reduced and optimized. For example, on the MMAct dataset, MoRA typically converges within 20 training epochs, fewer than that required by standard HAR models. Inference latency remains at the millisecond level, demonstrating MoRA’s efficiency and suitability for resource-constrained edge devices.

%% ============================================================================ %%

\section{Conclusion}
This paper introduces MoRA, a retrieval-augmented module for IMU-based HAR that addresses sample scarcity and limited knowledge utilization. Seamlessly integrable into existing architectures with minimal overhead, MoRA employs an uncertainty-adaptive fusion unit leveraging physical priors to reduce redundancy and enhance robustness. Experiments on ten benchmark datasets show consistent gains across diverse models and tasks, demonstrating strong generalizability.

%% ============================================================================ %%

\section*{Acknowledgment}

This work was supported by the Science and Technology Development Fund of the Macau SAR (Grant No. 0119/2025/RIA2) and the Guangdong Basic and Applied Basic Research Foundation (Grant No. 2025A1515050001).

% The preferred spelling of the word ``acknowledgment'' in America is without 
% an ``e'' after the ``g''. Avoid the stilted expression ``one of us (R. B. 
% G.) thanks $\ldots$''. Instead, try ``R. B. G. thanks$\ldots$''. Put sponsor 
% acknowledgments in the unnumbered footnote on the first page.

% \section*{References}

% Please number citations consecutively within brackets \cite{b1}. The 
% sentence punctuation follows the bracket \cite{b2}. Refer simply to the reference 
% number, as in \cite{b3}---do not use ``Ref. \cite{b3}'' or ``reference \cite{b3}'' except at 
% the beginning of a sentence: ``Reference \cite{b3} was the first $\ldots$''

% Number footnotes separately in superscripts. Place the actual footnote at 
% the bottom of the column in which it was cited. Do not put footnotes in the 
% abstract or reference list. Use letters for table footnotes.

% Unless there are six authors or more give all authors' names; do not use 
% ``et al.''. Papers that have not been published, even if they have been 
% submitted for publication, should be cited as ``unpublished'' \cite{b4}. Papers 
% that have been accepted for publication should be cited as ``in press'' \cite{b5}. 
% Capitalize only the first word in a paper title, except for proper nouns and 
% element symbols.

% For papers published in translation journals, please give the English 
% citation first, followed by the original foreign-language citation \cite{b6}.

\bibliographystyle{IEEEbib}
\bibliography{icme_references}

% \vspace{12pt}
% \color{red}
% IEEE conference templates contain guidance text for composing and formatting conference papers. Please ensure that all template text is removed from your conference paper prior to submission to the conference. Failure to remove the template text from your paper may result in your paper not being published.

\newpage

\appendix

\subsection{Related Work}

We briefly discuss two main lines of related work as follows.%: human activity recognition and retrieval-augmented modeling.

\noindent \textbf{Human Activity Recognition:}
With the rapid advancement of ubiquitous computing, HAR based on consumer smart devices has become a critical technology. Embedded IMUs continuously capture tri-axial acceleration and angular velocity, producing structured time-series data that enables fine-grained behavioral perception. In recent years, IMU-based HAR systems have transitioned from traditional signal processing methods using hand-crafted features to end-to-end learning frameworks~\cite{yue2022ts2vec, eldele2024tslanet} with convolutional neural networks and transformers, significantly improving recognition performance. To address performance degradation caused by data heterogeneity, researchers~\cite{xu2023practically} have explored strategies such as domain adaptation and personalized fine-tuning. 
% For instance, UniHAR~\cite{xu2023practically} uses the physical properties of IMU data to develop data augmentation techniques that integrate labeled and unlabeled samples, effectively addressing performance degradation induced by data shift. 
Although these methods yield promising results in specific scenarios, they fall short of meeting the growing demands of sensing systems for generalization and semantic understanding. 
With the advent of large language models, researchers have explored pre-training paradigms~\cite{moon2022imu2clip, girdhar2023imagebind, zhang2024unimts, han2024onellm} for IMU-based systems, significantly advancing zero-shot recognition and task transfer. 
% For example, OneLLM~\cite{han2024onellm} introduces a unified framework that aligns multiple modalities, including IMU data, with language, providing novel insights into multimodal semantic understanding. 
Although existing work has improved contextual reasoning and generalization, real-world deployment remains challenging. Large-scale collection and annotation of personalized data are costly and insufficient to cover real-world scenarios. Additionally, current methods inefficiently leverage transferable knowledge from historical data, lacking mechanisms for dynamically retrieving relevant information during inference, which limits their full potential.

\noindent \textbf{Retrieval-Augmented Modeling:}
Retrieval-augmented modeling is a general paradigm that uses explicit retrieval mechanisms to extract relevant information from knowledge bases or historical data dynamically. This approach has shown broad applicability and effectiveness in improving model reasoning in complex tasks. 
% For instance, in natural language processing~\cite{fan2024survey}, retrieval-augmented techniques have been employed to improve contextual consistency and factual accuracy in tasks like text generation and question answering. 
% In image generation, Remodiffuse~\cite{zhang2023remodiffuse} enhances visual fidelity and detail quality by retrieving similar images or local regions. In scientific reasoning, MolRAG~\cite{anonymous2025molrag} leverages structurally similar molecules as contextual references to improve molecular property prediction. 
Despite these advances, retrieval-augmented modeling has yet to be systematically explored in IMU-based HAR. Direct application of this paradigm~\cite{liu2024retrieval} faces two critical challenges. First, models must balance retrieval efficiency and semantic discriminability. In resource-constrained edge environments, models must construct compact yet semantically expressive embedding spaces to support fast and accurate approximate retrieval. Second, models must consider temporal dynamics and generalization. IMU data are inherently temporal, highly individualized, and susceptible to distribution shifts caused by variations in devices and environments. To address these challenges, we propose a universal retrieval-augmented module that enhances HAR model adaptability to complex behaviors and cross-user scenarios without increasing inference latency.

\subsection{Procedure of MoRA}
\label{sec:mora_code}

Algorithm 1 presents the overall procedure of MoRA in a self-contained and interpretable form.

\begin{algorithm}[htbp]
\LinesNotNumbered
\SetAlgoLined
\caption{Deployment Procedure of MoRA}
\KwIn{Given a test instance $x_q$, a historical dataset $\mathcal{D}_{\text{hist}} = {(x_j, y_j)}_{j=1}^{m}$
% ~\SL{Should be consistent to that in the main body of the paper}
, a fusion ratio $\alpha$, a pre-trained encoder $\mathcal{F}$ and a human activity recognition model $f_\theta$.}
\KwOut{Final prediction $\mathcal{C}_{\text{final}}(y \mid x_q)$.}
\BlankLine
\tcp{Building databases}
Initialize empty database $\mathcal{B}$.\;

$\mathcal{B} = \{(\mathcal{F}(x_j), y_j) \mid (x_j, y_j) \in \mathcal{D}_{\text{hist}}\}$

\tcp{Retrieving knowledge}
\For{each test instance $x_q$}{

    \tcp{Initial distribution}
    $\mathcal{C}_{\text{ref}}(y \mid x_q) = f_\theta (x_q)$\;
    
    \tcp{Retrieval-augmented distribution}
    Compute embedding $z_q = \mathcal{F}(x_q)$.\;
    
    Retrieve top-$k$ neighbors $\{y^{\text{db}}_1,...,y^{\text{db}}_k\}$ from $\mathcal{B}$.\;

    Compute semantic distribution $\mathcal{C}_{\text{rag}}(y \mid x_q, \mathcal{B})$.\;
    
    \eIf{dynamic fusion is required}{
    \tcp{Train fusion unit [UAFU]}
        $\alpha^* = \texttt{UAFU}(x_q)$ \tcp*{Adaptive ratio}
    }{
       $\alpha^* = \alpha$ \tcp*{Shared ratio}
    }
    \tcp{Final enhanced prediction}
    $\mathcal{C}_{\text{final}}(y \mid x_q) = \alpha^* \cdot \mathcal{C}_{\text{ref}}(y \mid x_q) + (1 - \alpha^*) \cdot \mathcal{C}_{\text{rag}}(y \mid x_q, \mathcal{B})$
}
\end{algorithm}

\subsection{Features}
\label{feature}

% We extract handcrafted features from five aspects: inter-channel correlation captures linear dependency among channels; global variance reflects the overall energy distribution; mean absolute deviation within channels reflects short-term dynamic changes; zero-crossing rate indicates signal oscillation and directional changes; and the range of local maximum variance highlights transitions between stable and volatile states. These features offer strong interpretability and efficiency while providing complementary discriminative information to the semantic embeddings.

We extract handcrafted features from five aspects, tailored to the HAR task. Specifically, inter-channel correlation captures linear dependencies across sensor channels, which is crucial for recognizing coordinated body movements. Global variance reflects the overall energy distribution of signals, offering insight into activity intensity. Mean absolute deviation within channels characterizes short-term dynamic variations, aiding in distinguishing fine-grained motion patterns. Zero-crossing rate indicates oscillatory behavior and directional changes, which are particularly relevant for periodic or transitional activities. Finally, the range of local maximum variance highlights shifts between stable and volatile states, capturing transitions inherent in complex actions. These features are selected for their strong interpretability and computational efficiency, and they provide complementary discriminative cues to semantic embeddings, thereby enhancing HAR performance.

\subsection{Loss Function}
\label{loss}

This subsection details the loss functions employed during the training of MoRA and their corresponding formulations. The overall loss consists of three components: the classification loss ($L_{\text{cls}}$), the alignment loss ($L_{\text{align}}$), and the sparsity loss ($L_{\text{sparse}}$), which respectively optimize the discriminative power, fusion rationality, and interpretability of the retrieval-augmented mechanism.

\textbf{The classification loss} ($L_{\text{cls}}$) measures the discrepancy between model predictions and ground-truth labels. We adopt the standard cross-entropy loss, defined as:
$$L_{cls} = - \sum_{i} y_i \log(\hat{y}_i),$$
where $y_i$ denotes the one-hot encoding of the true label, and $\hat{y}_i$ represents the predicted probability for the corresponding class, derived from the final prediction output $\mathcal{C}_{\text{final}}$. 

\textbf{The alignment loss} ($L_{\text{align}}$) is designed to guide and regularize the fusion weight $\alpha$ output by the MoRA module, enabling it to adaptively respond to the model's predictive uncertainty. Specifically, we first compute the entropy $H$ of the prediction distribution from the base IMU model to quantify its predictive uncertainty. 

Based on the computed entropy $H$, we construct a supervised target $\hat{\alpha}$ for the fusion weight, which is normalized into the range $[0,1]$. We adopt the binary cross-entropy loss to measure the discrepancy between the output fusion weight $\alpha$ from the gating network and its target $\hat{\alpha}$:
$$
L_{align} = - \frac{1}{N} \sum_{j=1}^{N} [\hat{\alpha}_j \log(\alpha_j) + (1 - \hat{\alpha}_j) \log(1 - \alpha_j)].
$$
By minimizing $L_{\text{align}}$, MoRA learns to perceive the confidence level of the base model and selectively incorporates retrieval information when uncertainty is high, thereby improving the overall reliability of decision-making.

\textbf{The sparsity loss} ($L_{\text{sparse}}$) aims to discourage the fusion weight $\alpha$ from concentrating around neutral values (e.g., 0.5), encouraging the model to make clearer decisions between relying on raw features or retrieval information and avoiding ambiguous fusion strategies. To enhance class-level adaptability, we introduce a learnable parameter $\beta$ to dynamically weight the contribution of different classes in the sparsity loss. It is defined as:
$$L_{sparse} = \text{mean}(\beta \times \alpha \times (1 - \alpha)).$$
This function reaches its minimum when $\alpha$ approaches 0 or 1, and its maximum around $\alpha = 0.5$, thereby encouraging the fusion weight to converge towards more decisive values.

Finally, the total loss for MoRA is a weighted combination of the above three components. By jointly minimizing $L_{\text{cls}}$, $L_{\text{align}}$, and $L_{\text{sparse}}$, the model not only enhances overall recognition performance but also effectively regulates the gating strategy, thereby ensuring greater robustness and interpretability in the knowledge fusion retrieval process.

\subsection{Datasets}
\label{dataset}

We conduct experiments on multiple real-world downstream evaluation datasets. The following subsection provides a brief overview of the sources and characteristics of each dataset:
\begin{itemize}
    \item UCI-HAR~\cite{anguita2013public}: Collected using a smartphone worn on the waist, this dataset contains six basic daily activities such as walking upstairs and downstairs.
    \item MotionSense~\cite{malekzadeh2019mobile}: Captured using a smartphone placed in the front trouser pocket, this dataset records six activities, including sitting and jogging.
    \item Shoaib~\cite{shoaib2014fusion}: Comprising seven daily activities (e.g., biking), this dataset was collected using sensors placed at five locations: left and right trouser pockets, belt, right upper arm, and right wrist.
    \item RealWorld~\cite{sztyler2016body}: Covers eight typical daily activities, with sensors placed on multiple body parts, including the chest, forearm, head, lower leg, and thigh.
    \item PAMAP~\cite{reiss2012introducing}: Captures twelve activities such as ironing, vacuuming, and skipping rope using sensors worn on the wrist, chest, and ankle.
    \item USC-HAD~\cite{zhang2012usc}: Contains twelve activity classes (e.g., sleeping, taking the elevator) recorded using a sensor fixed to the front of the right hip.
    \item WISDM~\cite{weiss2019wisdm}: Consists of 18 activities such as brushing teeth, drinking soup, and playing ball, recorded using a smartphone placed in a trouser pocket and a smart watch on the wrist.
    \item DSADS~\cite{altun2010comparative}: Encompasses 18 daily and sports activities (e.g., exercising, rowing), with sensors deployed on five body locations, including the torso and limbs.
    \item UTD-MHAD~\cite{chen2015utd}: Comprises 27 fine-grained actions (e.g., throwing, clapping, drawing), with sensors worn on the right wrist and right thigh.
    \item MMAct~\cite{kong2019mmact}: A large-scale and diverse human activity recognition dataset involving 35 daily behaviors (e.g., carrying objects, falling, making phone calls), collected via smartwatches and smartphones placed in trouser pockets.
\end{itemize}
The original sampling rates of these datasets mostly range from 50 to 100Hz. To ensure consistent processing and strike a balance between performance and computational efficiency, all datasets are downsampled to 20Hz.

\subsection{Baselines}
\label{baseline}

\begin{table*}[htbp]
  \centering
  \caption{Inference with fine-tuning [F1]. The best results are in \textbf{bold}, and the second-best results are \underline{underlined}.}
  \label{tab:full_fine-tuning}

  % 调整表格宽度
  \begin{adjustbox}{max width=\textwidth}
  \renewcommand{\arraystretch}{1.3}
    \begin{tabular}{c|c|ccccc}
      \hline
      \multirow{2}{*}{Methods} & \multirow{2}{*}{N-shot} & UCI-HAR & MotionSense & Shoaib & RealWorld & PAMAP \\
      \cline{3-7}
      & & ORG / MoRA[B] / MoRA[F] & ORG / MoRA[B] / MoRA[F] & ORG / MoRA[B] / MoRA[F] & ORG / MoRA[B] / MoRA[F] & ORG / MoRA[B] / MoRA[F] \\
      \hline\hline
      \multirow{6}{*}{Mantis} & 1 & 60.81 / \underline{62.33} / \textbf{63.84} & 53.79 / \underline{53.89} / \textbf{55.49} & 32.19 / \underline{32.84} / \textbf{33.66} & 30.97 / \underline{30.98} / \textbf{31.08} & 17.21 / \underline{17.46} / \textbf{18.50} \\
      \cline{2-7}
      & 2 & 67.09 / \underline{67.57} / \textbf{67.75} & 55.63 / \underline{55.56} / \textbf{55.67} & 51.35 / \underline{52.04} / \textbf{52.57} & 30.43 / \underline{30.46} / \textbf{30.66} & 32.42 / \underline{32.89} / \textbf{32.97} \\
      \cline{2-7}
      & 3 & 69.71 / \underline{70.06} / \textbf{70.52} & \underline{67.63} / 66.57 / \textbf{66.27} & 68.06 / \underline{68.43} / \textbf{68.49} & 31.29 / \underline{31.34} / \textbf{31.54} & 33.24 / \underline{33.55} / \textbf{33.96} \\
      \cline{2-7}
      & 5 & \textbf{71.92} / 71.43 / \underline{71.63} & 75.62 / \underline{75.65} / \textbf{75.72} & 63.72 / \underline{64.41} / \textbf{64.73} & 43.79 / \underline{43.91} / \textbf{44.05} & 46.22 / \underline{46.24} / \textbf{46.40} \\
      \cline{2-7}
      & 10 & 80.55 / \underline{81.01} / \textbf{81.77} & \textbf{89.17} / \underline{89.13} / 88.79 & 80.43 / \underline{80.89} / \textbf{82.16} & 45.79 / \underline{46.18} / \textbf{46.86} & \textbf{61.79} / \underline{61.32} / 60.63 \\
      \cline{2-7}
      & ALL & 94.70 / \underline{94.63} / \textbf{95.77} & 98.48 / \underline{98.61} / \textbf{99.31} & 96.60 / \underline{96.60} / \textbf{97.10} & 77.61 / \underline{78.01} / \textbf{78.99} & 82.27 / \underline{82.23} / \textbf{84.53} \\
      \hline\hline
      \multirow{6}{*}{UniMTS} & 1 & 50.30 / \underline{51.16} / \textbf{51.45} & 58.67 / \underline{58.95} / \textbf{59.02} & 73.36 / 72.85 / \underline{74.48} & 42.51 / \underline{42.92} / \textbf{43.54} & 59.44 / \underline{61.26} / \textbf{62.83} \\
      \cline{2-7}
      & 2 & 61.81 / \underline{62.68} / \textbf{62.78} & 77.50 / \underline{77.86} / \textbf{78.60} & 73.99 / \underline{77.03} / \textbf{77.81} & \textbf{71.61} / \underline{71.95} / 71.09 & 64.74 / \underline{64.94} / \textbf{65.29} \\
      \cline{2-7}
      & 3 & 71.79 / 71.55 / \underline{72.16} & 85.83 / 85.26 / \underline{85.61} & 81.46 / \underline{81.46} / \textbf{82.09} & 63.90 / 63.39 / \underline{64.06} & \textbf{80.39} / \underline{80.74} / 79.34 \\
      \cline{2-7}
      & 5 & \textbf{76.78} / 75.44 / \underline{76.18} & 82.09 / \underline{82.35} / \textbf{82.35} & 80.21 / \underline{80.55} / \textbf{81.11} & 68.87 / \underline{69.34} / \textbf{69.51} & 74.78 / \underline{75.30} / \textbf{75.31} \\
      \cline{2-7}
      & 10 & 78.93 / 78.64 / \underline{79.18} & 87.29 / \underline{87.74} / \textbf{87.82} & 89.72 / \underline{90.00} / \textbf{91.59} & 79.83 / \underline{80.32} / \textbf{80.89} & \textbf{77.20} / 75.43 / \underline{76.51} \\
      \cline{2-7}
      & ALL & 88.23 / \underline{88.51} / \textbf{90.48} & 95.14 / \underline{95.14} / \textbf{96.51} & 97.53 / \underline{97.84} / \textbf{98.68} & 84.91 / \underline{85.10} / \textbf{85.96} & \textbf{90.26} / \underline{90.11} / 90.10 \\
      \hline
      \multirow{2}{*}{Methods} & \multirow{2}{*}{N-shot} & USC-HAD & WISDM & DSADS & UTD-MHAD & MMAct \\
      \cline{3-7}
      & & ORG / MoRA[B] / MoRA[F] & ORG / MoRA[B] / MoRA[F] & ORG / MoRA[B] / MoRA[F] & ORG / MoRA[B] / MoRA[F] & ORG / MoRA[B] / MoRA[F] \\
      \hline\hline
      \multirow{6}{*}{Mantis} & 1 & 41.25 / \underline{41.56} / \textbf{41.59} & 38.51 / \underline{38.89} / \textbf{39.56} & 2.99 / \underline{3.02} / \textbf{3.07} & 11.08 / \underline{11.07} / \textbf{11.34} & 11.66 / \underline{11.70} / \textbf{11.71} \\
      \cline{2-7}
      & 2 & 47.72 / \underline{47.78} / \textbf{48.85} & 42.40 / \underline{42.70} / \textbf{43.32} & \textbf{5.05} / \underline{5.02} / 5.02 & 19.11 / \underline{19.09} / \textbf{19.15} & 12.05 / \underline{12.08} / \textbf{12.27} \\
      \cline{2-7}
      & 3 & 56.13 / \underline{56.47} / \textbf{56.96} & 56.90 / \underline{56.74} / \textbf{56.92} & 5.52 / \underline{5.56} / \textbf{5.65} & 27.58 / \underline{27.53} / \textbf{27.58} & 17.35 / \underline{17.31} / \textbf{17.57} \\
      \cline{2-7}
      & 5 & 52.71 / \underline{52.83} / \textbf{52.99} & \textbf{59.30} / \underline{59.17} / 58.76 & 5.67 / \underline{5.67} / \textbf{5.72} & 30.92 / \underline{31.18} / \textbf{31.24} & 19.14 / \underline{19.15} / \textbf{19.18} \\
      \cline{2-7}
      & 10 & 61.93 / \underline{62.16} / \textbf{63.58} & 63.61 / \underline{63.79} / \textbf{64.56} & 5.34 / \underline{5.37} / \textbf{5.49} & 45.23 / \underline{45.23} / \textbf{45.60} & 24.91 / \underline{24.92} / \textbf{25.04} \\
      \cline{2-7}
      & ALL & 74.36 / \underline{74.38} / \textbf{75.52} & 76.99 / \underline{77.05} / \textbf{78.36} & 79.56 / \underline{79.57} / \textbf{81.66} & 54.41 / \underline{54.08} / \textbf{55.69} & 49.55 / \underline{49.58} / \textbf{51.09} \\
      \hline\hline
      \multirow{6}{*}{UniMTS} & 1 & 39.06 / \underline{39.49} / \textbf{39.78} & 30.11 / \underline{30.27} / \textbf{30.47} & 62.93 / \underline{63.09} / \textbf{63.22} & 41.42 / \underline{41.41} / \textbf{43.33} & 13.21 / \underline{13.16} / \textbf{13.24} \\
      \cline{2-7}
      & 2 & 42.45 / \underline{42.20} / \textbf{42.52} & 42.46 / \underline{42.78} / \textbf{42.79} & 63.17 / \underline{63.07} / \textbf{63.96} & 50.06 / \underline{50.06} / \textbf{50.50} & 24.26 / \underline{24.41} / \textbf{24.42} \\
      \cline{2-7}
      & 3 & 44.81 / \underline{44.48} / \textbf{44.37} & 46.37 / \underline{46.65} / \textbf{46.88} & \textbf{69.72} / \underline{67.94} / 68.18 & 56.16 / \underline{56.97} / \textbf{58.48} & 25.83 / \underline{26.05} / \textbf{26.26} \\
      \cline{2-7}
      & 5 & 49.67 / \underline{49.91} / \textbf{50.59} & 52.30 / \underline{52.44} / \textbf{52.49} & \textbf{72.24} / \underline{71.12} / 70.76 & 59.66 / \underline{59.58} / \textbf{59.91} & 34.92 / \underline{34.95} / \textbf{34.97} \\
      \cline{2-7}
      & 10 & 53.73 / \underline{53.49} / \textbf{54.22} & 54.95 / \underline{55.13} / \textbf{55.24} & 79.60 / \underline{79.42} / \textbf{80.47} & 66.52 / \underline{66.41} / \textbf{67.09} & 37.88 / \underline{37.98} / \textbf{38.13} \\
      \cline{2-7}
      & ALL & 59.51 / \underline{59.45} / \textbf{59.81} & 71.86 / \underline{71.81} / \textbf{73.71} & 92.67 / \underline{92.62} / \textbf{93.76} & 68.30 / \underline{68.44} / \textbf{70.09} & 55.33 / \underline{55.33} / \textbf{56.28} \\
      \hline
    \end{tabular}
  \end{adjustbox}
\end{table*}

While all selected baseline methods are based on open-source implementations, several SOTA approaches, such as TS2Vec, UniHAR, TSLANet, TimeMixer, and IMU2CLIP do not provide publicly available pretrained weights. Ego4D~\cite{grauman2024ego} is one of the largest first-person multimodal activity recognition datasets to date, comprising IMU, visual, and textual modalities, and thus serves as a strong foundation for high-quality pretraining. The deployment details of each method are as follows:
\begin{itemize}
    \item TS2Vec~\cite{yue2022ts2vec}: Utilizes a time-series framework to obtain embeddings. Following the original paper, we first train the encoder using contrastive learning, then train a linear classifier on each downstream dataset.
    \item UniHAR~\cite{xu2023practically}: Trains the encoder via contrastive learning based on the data augmentation strategy described in the original paper, followed by training a classifier on each dataset.
    \item TSLANet~\cite{eldele2024tslanet}: Employs a self-supervised learning framework based on time-series segmentation and positional embeddings, where the encoder is trained first, followed by a linear classifier.
    \item Mantis~\cite{feofanov2025mantis}: A time-series classification model based on a vision Transformer architecture. We utilize publicly available pretrained weights and append a linear layer to match the number of task-specific classes.
    \item TimeMixer~\cite{wang2024timemixer++}: Following the settings in the original paper, TimeMixer performs end-to-end training by decomposing temporal signals into trend and seasonal components to enable classification.
    \item IMU2CLIP~\cite{moon2022imu2clip}: Although pretrained on Ego4D, the original weights are unavailable. We retrain its encoder using the original configuration.
    \item UniMTS~\cite{zhang2024unimts}: Evaluated using the authors’ publicly released pretrained weights, with a linear mapping layer added during fine-tuning to support classification.
    \item PRIMUS~\cite{das2025primus}: Directly evaluated using the pretrained model provided by the authors.
    \item ImageBind~\cite{girdhar2023imagebind}: Performs zero-shot inference using the publicly available pretrained model.
    \item OneLLM~\cite{han2024onellm}: Utilizes public weights, vectorizes the IMU signals, and inputs them into a language model, which generates a response to construct a probability distribution over class labels, enabling zero-shot inference.
\end{itemize}
More implementation details will be made available through a public code repository upon acceptance of the paper.

\subsection{Evaluation}
\label{eva_1}

% \subsection{Overall Performance}
The complete experimental results of retrieval-augmented inference with fine-tuning are presented in Table~\ref{tab:full_fine-tuning}, which validate the effectiveness of MoRA. Furthermore, MoRA[F], the full version incorporating the uncertainty-adaptive fusion mechanism, demonstrates superior performance across various datasets and task settings, further confirming the effectiveness of the proposed fusion unit.

\begin{figure*}
    \centering
    \includegraphics[width=\textwidth]{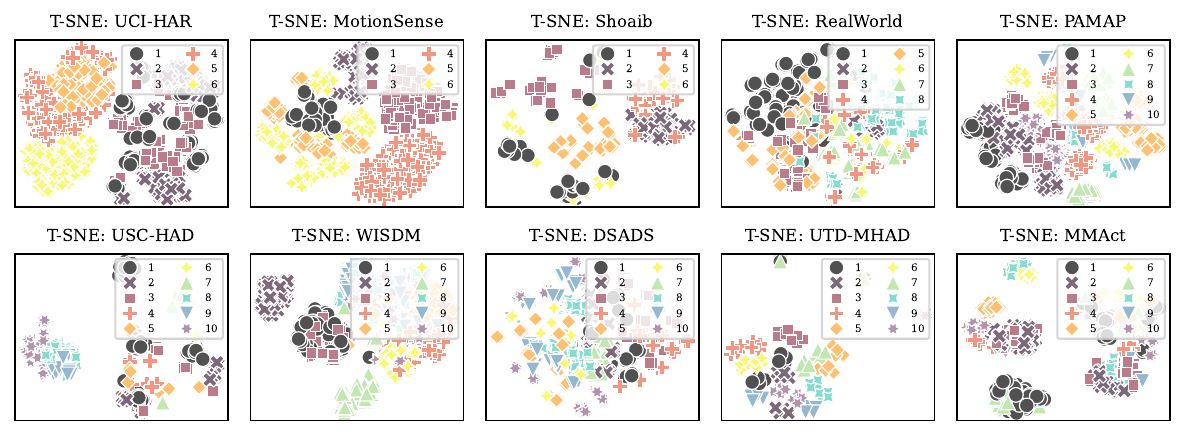}
    \caption{T-SNE-based feature visualization of representations learned by the Mantis model.}
    \label{fig:mantis_tsne}
\end{figure*}

\begin{figure*}
    \centering
    \includegraphics[width=\textwidth]{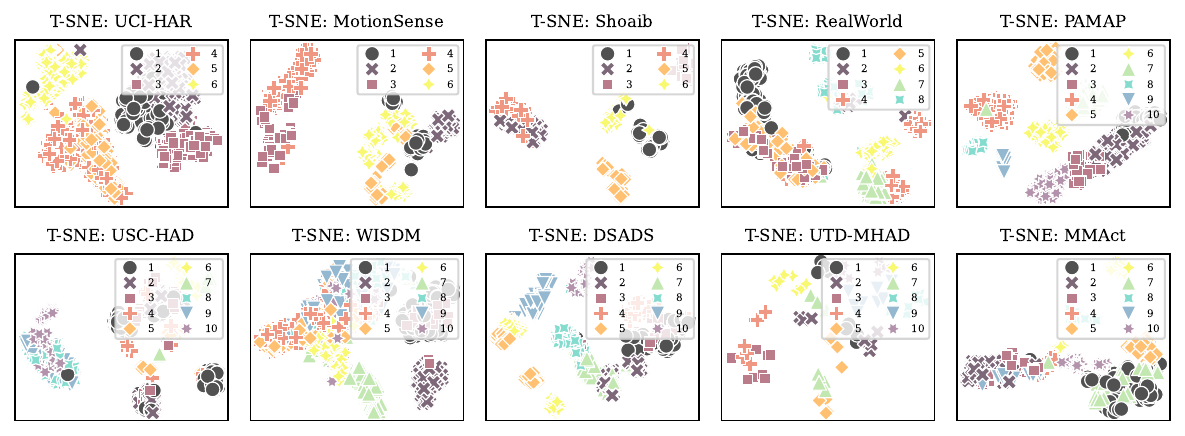}
    \caption{T-SNE-based feature visualization of representations learned by the UniMTS model.}
    \label{fig:unimts_tsne}
\end{figure*}

\begin{figure*}
    \centering
    \includegraphics[width=\textwidth]{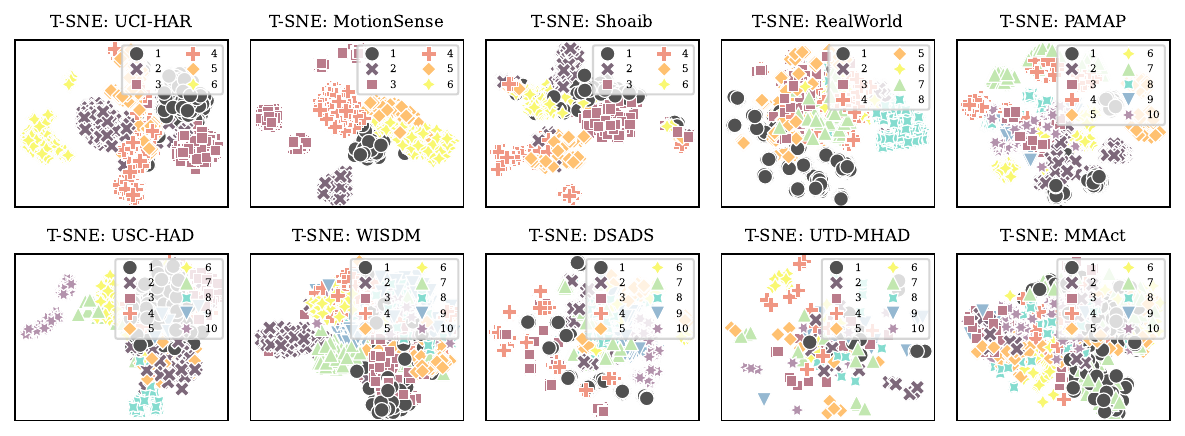}
    \caption{T-SNE-based feature visualization of representations learned by the TimeMixer model.}
    \label{fig:timemixer_tsne}
\end{figure*}

Figures~\ref{fig:mantis_tsne}, ~\ref{fig:unimts_tsne}, and ~\ref{fig:timemixer_tsne} illustrate the T-SNE visualizations of the embedding representations from the baseline models Mantis, UniMTS, and TimeMixer, respectively. Although Mantis demonstrates superiority, its embedding clustering quality remains suboptimal on certain datasets. The quality of embeddings is closely tied to the performance of the retrieval-augmented module. We need to explore more efficient and robust knowledge acquisition methods.

\end{document}